\documentclass{article}

\usepackage[height=22.5cm, width=16.5cm, hmarginratio={1:1}]{geometry}

\usepackage{tikz-cd}
\usepackage{amsmath}
\usepackage{amsfonts}
\usepackage{mathrsfs}
\usepackage[utf8]{inputenc}
\usepackage[maxbibnames=99]{biblatex}
\usepackage{url}

\renewbibmacro{in:}{}
\DeclareFieldFormat[article]{pages}{#1}
\DeclareFieldFormat[incollection]{pages}{#1}
\DeclareFieldFormat[inbook]{citetitle}{#1}
\DeclareFieldFormat[inbook]{title}{#1} 
\DeclareFieldFormat[article]{citetitle}{#1}
\DeclareFieldFormat[article]{title}{#1} 
\DeclareFieldFormat[incollection]{title}{#1} 
\DeclareFieldFormat[incollection]{booktitle}{#1} 
\DeclareFieldFormat{journaltitle}{#1\isdot}
\DeclareFieldFormat{title}{#1\isdot}
\DeclareFieldFormat[article]{volume}{\mkbibbold{#1}}
\DeclareNameAlias{default}{last-first}

\DefineBibliographyExtras{english}{%
}

\usepackage[thmmarks]{ntheorem}
{	
	\theoremstyle{nonumberplain}
	\theoremheaderfont{\bfseries}
	\theorembodyfont{\normalfont}
	\theoremsymbol{\mbox{$\Box$}}
	
}
\usepackage{indentfirst}
\newtheorem{prop}{Proposition}[section]

\newtheorem{exmp}{Example}[section]

\title{Several Special Solutions of Open WDVV Equations}

\AtEndDocument{\bigskip{\footnotesize%
  \textsc{School of Mathematical Sciences, University of Science and Technology of China, Hefei 230026, P. R. China} \par  
  \textit{E-mail address}: \texttt{fzsheng2000@mail.ustc.edu.cn} \par
}}

\usepackage{lipsum}

\begin{document}
\author{Fangze Sheng}
\date{}
\maketitle

\begin{abstract}
	The Witten-Dijkgraaf-Verlinde-Verlinde(WDVV) equations appeared in the study of two-dimensional topological field theoies in the early 1990s\cite{witten1990two}\cite{Dijkgraaf:1990dj}. An  extension of the WDVV equations, called the open WDVV equations, was introduced by A.Horev and J.P.Solomon\cite{HS12}. In this paper, we give some particular solutions to the open WDVV equations.
\end{abstract} 

\section{Introduction}
	The WDVV equations are the following systems of equations\footnote{These systems of equations are called WDVV associativity equations in Dubrovin's original literature \cite{Dub}. Besides the associativity equations, the WDVV equations in \cite{Dub} also contain homogeneous equations. For simplicity, we call the associativity part the WDVV equations in this paper.}:
	\begin{equation}\label{WDVV}
		\dfrac{\partial^3F}{\partial t^{\alpha}\partial t^{\beta}\partial t^{\mu}}\eta^{\mu\nu}\dfrac{\partial^3F}{\partial t^{\nu}\partial t^{\gamma}\partial t^{\delta}}=\dfrac{\partial^3F}{\partial t^{\delta}\partial t^{\beta}\partial t^{\mu}}\eta^{\mu\nu}\dfrac{\partial^3F}{\partial t^{\nu}\partial t^{\gamma}\partial t^{\alpha}}, 
	\end{equation}
	that hold for $1\leq \alpha,\beta,\gamma,\delta \leq N$. Here\ $F=F(t^1,\dots,t^N)$ is an analytic function defined on some open sets $M\subset\mathbb{C}^N$. $\eta=(\eta_{
	\alpha \beta})$ is an $N\times N$ dimensional constant non-degenerate matrix and $(\eta^{
	\alpha \beta}):=\eta^{-1}$. We use the convention of summing over repeated Greek indices. The corresponding Frobenius manifolds on $M$ can be defined due to the work of Dubrovin \cite{Dub}.
	\par Particular examples of solutions to WDVV equations are interesting. B.Dubrovin constructs closed Frobenius structures on the space of orbits of Coxeter groups, where all of solutions to WDVV equations are polynomials \cite{dubrovin1993differential}. B.Dubrovin gives the closed potential of extended affine group $A_1^{(1)}$ in \cite{Dub}. For a particular choice of basis of the root systems, B.Dubrovin and Y.Zhang construct Frobenius structures on the orbit spaces of the extended affine Weyl groups \cite{dubrovin1998extended}. In fact, the geometric structure revealed in \cite{dubrovin1998extended} does not depend on the choice of basis of root systems in case $A_{l}$, $B_{l}$, $C_{l}$ and $D_{l}$, according to the work of B.Dubrovin, Y.Zhang \cite{dubrovin1998extended}(case $A_{l}$) and B.Dubrovin, I.Strachan, Y.Zhang, D.Zuo \cite{Dubrovin_2019}(case $B_{l}$, $C_{l}$, $D_{l}$). In these cases, the solutions are triangular polynomials. In \cite{kontsevich1994gromov}, M.Kontsevich and Yu.I.Manin show that solutions of quantum cohomology type are triangular series.  

	\par Recently, another system of equations appears in the research of Gromov-Witten theory \cite{HS12}. Let $F=F(t^1,\dots,t^N)$ be a solution of the WDVV equations. The open WDVV equations associated to $F$ are the following system of equations, whose solution $F^{o}=F^{o}(t^1,\dots,t^N,s)$ depends on another variable $s$: 
	\begin{equation}
		\dfrac{\partial^3F}{\partial t^{\alpha}\partial t^{\beta}\partial t^{\mu}}\eta^{\mu\nu}\dfrac{\partial^2 F^{o}}{\partial t^{\nu}\partial t^{\gamma}}+\dfrac{\partial^2 F^{o}}{\partial t^{\alpha}\partial t^{\beta}}\dfrac{\partial^2 F^{o}}{\partial s\partial t^{\gamma}}=\dfrac{\partial^3F}{\partial t^{\gamma}\partial t^{\beta}\partial t^{\mu}}\eta^{\mu\nu}\dfrac{\partial^2 F^{o}}{\partial t^{\nu}\partial t^{\alpha}}+\dfrac{\partial^2 F^{o}}{\partial t^{\gamma}\partial t^{\beta}}\dfrac{\partial^2 F^{o}}{\partial s\partial t^{\alpha}},
	\end{equation}
	\begin{equation}
		\dfrac{\partial^3F}{\partial t^{\alpha}\partial t^{\beta}\partial t^{\mu}}\eta^{\mu\nu}\dfrac{\partial^2 F^{o}}{\partial t^{\nu}\partial s}+\dfrac{\partial^2 F^{o}}{\partial t^{\alpha}\partial t^{\beta}}\dfrac{\partial^2 F^{o}}{\partial s^2}=\dfrac{\partial^2 F^{o}}{\partial s\partial t^{\beta}}\dfrac{\partial^2 F^{o}}{\partial s\partial t^{\alpha}},
	\end{equation}
	that hold for $1\leq \alpha,\beta,\gamma,\delta \leq N$. 
	We consider solutions satisfying the following quasihomogenous condition \cite{BB20} 
	\begin{equation}
		E^{\gamma}\dfrac{\partial F^{o}}{\partial t^{\gamma}}+\dfrac{1-d}{2} s\dfrac{\partial F^{o}}{\partial s}=\dfrac{3-d}{2}F^{o}+D_{\gamma}t^{\gamma}+\widetilde{D}s+E,\quad \  D_{\gamma},\widetilde{D}, E\in \mathbb{C} 
	\end{equation}
	and additional conditions 
	\begin{equation}
		\ \dfrac{\partial^2 F^{o}}{\partial t^{1}\partial t^{\alpha}}=0,\ \dfrac{\partial^2 F^{o}}{\partial t^{1}\partial s}=1. \label{fujiatiaojian2}
	\end{equation}	
	
	A generalization of Frobenius manifolds, called F-manifolds, was introduced by C.Hertling and Y.Manin \cite{HertlingManin1999weak}. Flat F-manifolds were studied by E.Getzler\cite{getzler2004jet} and Y.Manin\cite{Man05}. According to A.Alcolado \cite{alcolado2017extended} and P.Rossi, open WDVV equations can be interpreted as the rank-1 extension of the Dubrovin-Frobenius manifold to flat F-manifold.

	\par It is natural to consider the solutions of open WDVV equations associated to the Dubrovin-Frobenius manifolds constructed from other finite irreducible Coxeter groups, extended affine Weyl groups and quantum cohomology.

 	A.Adam proves the existence of polynomial solutions to open WDVV equations associated to the extension of $A_N$ singularity \cite{alcolado2017extended}.  A.Basalaev and A.Buryak give the explicit solution in \cite{BB20}.
	In fact, more generally, for finite irreducible Coxeter groups $A_N$, $B_N$, $D_N$ and $I_2(k), k\geq3$, A.Basalaev and A.Buryak also give the following explicit solutions to the associated open WDVV equations: \cite{BB20}	\cite{B21}
	\par $A_N$: $F^o_{A_N}=F^o_{A_N}(t^1,...,t^N,s)$, with 
	\begin{equation*}\nonumber
		\dfrac{\partial^{m+k}F^o_{A_N}}{\partial t^{\alpha _1}...\partial t^{\alpha _m}(\partial s)^k}\bigg|_{t^1 =...= t^N = s = 0} = 	\left\{
			\begin{array}{ll}
  			(m+k-2)!\text{, if\ }\sum^m_{i=1}(N+2-\alpha _i)+k = N+2 \text{,} 
  			\\ 0, \quad \text{ otherwise.}
  			\end{array}\right.
	\end{equation*}

	$B_N$:  $F^o_{B_N}(t^1,...,t^N,s):=F^o_{A_{2N-1}}(t^1,0,t^2,0,...,t^{N-1},0,t^N,s)$.

	$D_N$:  $F^o_{D_N}(t^1,...,t^N,s):=\dfrac{s^{2N-1}}{2^{N-2}(2N-1)(2N-2)}+\dfrac{(t^N)^2}{2s}+\sum^{N-1}_{k=1}\dfrac{v^D_k s^{2k-1}}{2^{k-1}(2k-1)} $, with

	\begin{equation*} 
		v^D_b = \sum_{\alpha_1,...,\alpha_{N-1}\geq 0, \newline \sum^{N-1}_{k=1}(N-k)\alpha_k=N-b}\dfrac{(|\alpha|+2b-3)!}{(2b-2)!}\prod^{N-1}_{k=1}\dfrac{(t^k)^{\alpha_k}}{\alpha_k!}.
	\end{equation*}
		
	$I_2(k-1)$: 	
	\begin{equation*}
		F^o_{I_2(k-1)}(t^1,t^2,s) = 	\left\{
			\begin{array}{ll}
			st^1+\sum^{q-1}_{m=0} a_m s^{2q-2m}(t^2)^m,\quad k=2q, \label{closed}
  			\\ st^1+\sum^{q}_{m=0}b_m s^{2q-2m+1}(t^2)^m,\quad  k=2q+1. 
  			\end{array}\right.
	\end{equation*}
 Here, let $a_q=1$, then $a_n$ satisfies the following recursion relations \cite{BB20}
 	\begin{equation*}
	\begin{aligned}
 	a_{q-1}&=2(2q-1), 
	\\ a_{s-q}&=\dfrac{\sum^{q-1}_{m=s-q+1}m(2q-2s+2m)(m(4q-3)-(1+s)2q+2s+1)a_m a_{s-m}}{(4q-2s)(4q-2s-1)q(q-1)}\\&\ q+1\leq s\leq 2q-2. 
	\end{aligned}	
	 \end{equation*}
    $b_n$ satisfies \cite{BB20}
	\begin{equation*} 
	\begin{aligned}
	 b_q^2&=\dfrac{2(2q+1)(2q-1)}{q}, 
	 \\	b_{s-q}&=-\dfrac{\sum_{m=s-q+1}^{q-1}m(1+2m+2q-2s)(m(-1+4q)+s-2q(1+s))b_m b_{s-m}}{2q^2(1+4q-2s)(-1+2q-s)b_q}	, \\&q\leq s\leq 2q-1. 
	\end{aligned}
	\end{equation*}
 
We give a closed formula for $a_m$:
	\begin{equation}
		a_m=\dfrac{2^{2m+1}}{(2m+2)!} \dfrac{(2q-1)^m(q-m+1)_{m-2}(q+1)_{m-3}}{(q-1)^mq^m},\quad 0\leq m\leq q-1, 
    \end{equation}
where $(s)_n=s(s+1)\dots(s+n-1)$. 

    For the construction of type $A_N$ open invariants of higher genera, see \cite{BY15}\cite{Bur16}\cite{BCT21}\cite{buryak2018open}\cite{buryak2023open}\cite{pandharipande2022intersectiontheorymodulidisks}. 

	\par 
	For the extended affine Weyl group case $A_1^{(1)}$, A.Basalaev and A.Buryak obtain the following open solutions \cite{BB19}
	\begin{equation*}
		F^o_{A_1^{(1)}}(t^1,t^2,s)=t^1 s\pm 2\alpha^{-1}e^{\frac{t^2}{2}}\text{sinh}(\alpha(s+\beta)),\quad \alpha,\beta \in \mathbb{C}.
	\end{equation*}
	
	\par 
	In this paper, we obtain two new solutions to open WDVV equations, which are associated to the Coxeter group $H_3$ and the extended affine Weyl group $A_2^{(1)}$.

\section{Open solutions related to Coxeter groups $H_3$ and extended affine Weyl groups $A_2^{(1)}$}
	In the following section, we give two solutions to open WDVV equations associated to the Dubrovin-Frobenius manifolds of the Coxeter group $H_3$ and the extended affine Weyl group $A_2^{(1)}$. They are new as far as we know.
\subsection{$H_3$ case}
 	The Dubrovin-Frobenius potential for Coxeter group $H_3$ \cite{Dub} is 
 \begin{equation}
 	F^c_{H_3}(t^1,t^2,t^3)=\dfrac{1}{2}(t^1)^2t^3+\dfrac{1}{2}t^1(t^2)^2+\dfrac{1}{6}(t^2)^3(t^3)^2+\dfrac{1}{20}(t^2)^2(t^3)^5+\dfrac{1}{3960}(t^3)^{11}.
 \end{equation}
In this case, the constant $d=\dfrac{4}{5}$ and we consider solutions $F^o$ satisfying the quasihomogeneous equation  
   \begin{equation}
	t^1 \dfrac{\partial F^{o}}{\partial t^1}+\dfrac{3}{5}t^2 \dfrac{\partial F^{o}}{\partial t^2}+\dfrac{1}{5}t^3 \dfrac{\partial F^{o}}{\partial t^3}+\dfrac{1}{10} s\dfrac{\partial F^{o}}{\partial s} = \dfrac{11}{10} F^{o} .
  \end{equation}

\begin{prop}
	A solution of the open WDVV equations for $H_3$ is
 \begin{equation}
 	F^o_{H_3}(t^1,t^2,t^3,s)=s t^1+\frac{(t^2)^2}{2 s}+t^2 \left(\frac{(t^3)^2 s}{2}-\frac{t^3 s^3}{4}+\frac{s^5}{40}\right)-\frac{(t^3)^2 s^7}{32}+\frac{(t^3)^3 s^5}{8}-\frac{5 (t^3)^4 s^3}{24}+\frac{(t^3)^5 s}{10}+\frac{t^3 s^9}{288}-\frac{s^{11}}{7040}.	
 \end{equation}
Note that it has a simple pole at $s=0$, which is similar to the case $D_N$. 
\end{prop}

\subsection{$A_2^{(1)}$ case}
	The Dubrovin-Frobenius potential for extended affine Weyl group $A_2^{(1)}$ \cite{Dub} is 
\begin{equation}
	F^c_{A_2^{(1)}}=\dfrac{1}{2}(t^1)^2t^3+\dfrac{1}{2}t^1(t^2)^2-\dfrac{1}{24}(t^2)^4+t^2 e^{t^3}.	
\end{equation}
In this case, the constant $d=1$ and we consider solutions $F^o$ satisfying the quasihomogeneous equation  
  \begin{equation}
	 t^1 \dfrac{\partial F^{o}}{\partial t^1}+\dfrac{1}{2}t^2 \dfrac{\partial F^{o}}{\partial t^2}+\dfrac{3}{2}  \dfrac{\partial F^{o}}{\partial t^3} = F^{o}.	
  \end{equation}

 Using the methods of characteristics, we obtain the following solution to $A_2^{(1)}$:
\begin{prop}
	A solution of the open WDVV equations for $A_2^{(1)}$ is
	\begin{equation}
		F^o_{A_2^{(1)}}(t^1,t^2,t^3,s)=s t^1+(t^2)^2 \sum_{p,q\geq 0}a_{p,q}s^p (t^2-3 {\rm log}(t^3))^q,
	\end{equation}
	
	\begin{equation}
	a_{p,q}=\left\{
			\begin{array}{ll}
  				\frac{(-1)^{\frac{p-1}{2}} 2^{2 p-6} \sum _{j=1}^{\frac{p+1}{2}} j^q \left(\sum 	_{i=1}^j i^{-2 j+p+3} b_{i,j}\right)}{p! q!},\quad \text{p is odd} ,
  			\\ \frac{(-1)^{p/2} \left(\sum _{j=2}^{\frac{p}{2}+1} (2 j-1)^q \left(\sum _{i=1}^j (-1)^{i+1} (4 i-2)^{-2 j+p+4} d_{i,j}\right)+2^{p-1}\right)}{2^{q-1} p! q!},\quad \text{p is even}.		
  			\end{array}\right. 	\label{n=4OWDVV}	
 	\end{equation}
  	The coefficients\ $b_{i,j},d_{i,j}$ satisfy the following recursion relations: 
 	\begin{equation}
  		b_{1,1}=-8,\ d_{1,1}=1.
  	\end{equation}
  
	For $p=2k$, the relation is   
	\begin{equation}
	\begin{aligned}
	\tilde{A}_{k,q}+&\sum _{j=0}^q \sum _{r=2}^{k+1} \sum _{m=1}^r \tilde{B}_{k,q,m,r,j} d_{m,r}+\sum _{i=0}^k \sum _{j=0}^q \sum _{t=1}^{-i+k+1} \sum _{s=1}^t \tilde{C}_{k,q,s,t,i,j} b_{s,t}
    \\&+\sum _{i=0}^k \sum _{j=0}^q \sum _{t=1}^i \sum _{s=1}^t \tilde{D}_{k,q,s,t,i,j} b_{s,t}+\sum _{i=0}^k \sum _{j=0}^q \sum _{t=1}^{-i+k+1} \sum _{s=1}^t \sum _{r=2}^{i+1} \sum _{m=1}^r \tilde{E}_{k,q,s,t,m,r,i,j} b_{s,t} d_{m,r}
    \\&\quad +\sum _{i=0}^k \sum _{j=0}^q \sum _{t=1}^i \sum _{s=1}^t \sum _{r=2}^{-i+k+2} \sum _{m=1}^r \tilde{F}_{k,q,s,t,m,r,i,j} b_{s,t} d_{m,r}=0.
	\end{aligned}
	\end{equation}
	
	For $p=2k+1$, the relation is
	\begin{equation}
	\begin{aligned}
		\sum _{j=0}^q \sum _{r=1}^{k+1}& \sum _{m=1}^r \tilde{G}_{k,q,m,r} b_{m,r}+\sum _{i=0}^k \sum _{j=0}^q \sum _{r=2}^{i+1} \sum _{m=1}^r \tilde{H}_{k,q,m,r} d_{m,r}
        \\&+\sum _{i=0}^k \sum _{j=0}^q \sum _{t=2}^{-i+k+2} \sum _{s=1}^t \tilde{I}_{k,q,s,t} d_{s,t}+\sum _{i=0}^k \sum _{j=0}^q \sum _{r=1}^{i+1} \sum _{t=1}^{-i+k+1} \sum _{m=1}^r \sum _{s=1}^t \tilde{J}_{k,q,m,r,s,t} b_{m,r} b_{s,t}
        \\&\qquad+\sum _{i=0}^k \sum _{j=0}^q \sum _{r=2}^{i+1} \sum _{t=2}^{-i+k+2} \sum _{m=1}^r \sum _{s=1}^t \tilde{K}_{k,q,m,r,s,t} d_{m,r} d_{s,t}=0	,		
	\end{aligned}
	\end{equation}	
with coefficients given by
	\begin{equation}
		\tilde{A}_{k,q}=\sum _{j=0}^q \frac{(-1)^k 2^{-j+2 k-2}}{j! (2 k)! (q-j)!},
	\end{equation}
 
	\begin{equation}
 		\tilde{B}_{k,q,m,r,j}=\frac{\left(36 r^2-72 r+35\right) (2 r-1)^j (-1)^{k+m} 2^{-j+2 k-2 r+3} (2 m-1)^{2 k-2 r+4}}{j! (2 k)! (q-j)!},
	\end{equation}
	
	\begin{equation}
		\tilde{C}_{k,q,s,t,i,j}=\frac{ (-1)^{k+1} (1-2 t) (-2 i+2 k+1) 2^{-2 i-j+4 k-5} t^{q-j} s^{-2 (i-k+t-2)}}{(2 i)! (j)! (-2 i+2 k+1)! (q-j)!},
	\end{equation}
	
	\begin{equation}
		\tilde{D}_{k,q,s,t,i,j}=\frac{(-1)^{k+1} (2 t-1) (-i+k+1) t^{j+1} s^{2 i-2 t+2} 2^{2 i+j+2 k-q-5}}{(2 i-1)! j! (-2 i+2 k+2)! (q-j)!},
	\end{equation}
	
	\begin{equation}
    \begin{aligned}
		&\tilde{E}_{k,q,s,t,m,r,i,j}=
    \\&\qquad \frac{(-1)^{k+m} (2 r-2 t-1)(-2 i+2 k+1) (2 r-1)^{j+1}  (2 m-1)^{2 i-2 r+4} 2^{-2 i-j+4 k-2 r} s^{-2 (i-k+t-2)}t^{q-j}}{(2 i)! j! (-2 i+2 k+1)! (q-j)!},
    \end{aligned}
	\end{equation}

	\begin{equation}
    \begin{aligned}
     	&\tilde{F}_{k,q,s,t,m,r,i,j}=
     \\&\qquad \frac{(-1)^{k+m} (-2 r+2 t+1)(-i+k+1) (2 r-1)^{q-j}(2 m-1)^{-2 i+2 k-2 r+6} 2^{2 i+j+2 k-q-2 r} s^{2 i-2 t+2}t^{j+1} }{(2 i-1)! j! (-2 i+2 k+2)! (q-j)!},
    \end{aligned}
	\end{equation}

	\begin{equation}
		\tilde{G}_{k,q,m,r}=\frac{(-1)^{k+1} 16^{k-1} \left(9 r^2-9 r+2\right) r^j m^{2 k-2 r+4}}{j! (2 k+1)! (q-j)!},
	\end{equation}
	
	\begin{equation}
		\tilde{H}_{k,q,m,r}=\frac{(-1)^{k+m+1}2^{2 k-q-2 r+8}(r-1) (-i+k+1) (2 r-1)^{j+1} (2 m-1)^{2 i-2 r+4} }{(2 i)! j! (-2 i+2 k+2)! (q-j)!}	,\end{equation}

	\begin{equation}
		\tilde{I}_{k,q,s,t}=\frac{(-1)^{k+s+1} 2^{2 k-q-2 t+8}(1-t) (-i+k+1) (2 t-1)^{q-j}(2 s-1)^{-2 i+2 k-2 t+6}  }{(2 i)! j! (-2 i+2 k+2)! (q-j)!},	\end{equation}
	
	\begin{equation}
		\tilde{J}_{k,q,m,r,s,t}=\frac{(-1)^{k+1} 2^{4 k-7}(r-t) r^{j+1}  m^{2 i-2 r+4} t^{q-j} s^{-2 (i-k+t-2)}}{(2 i+1)! j! (2 k-2 i)! (q-j)!},
	\end{equation}
	
	\begin{equation}
    \begin{aligned}
		&\tilde{K}_{k,q,m,r,s,t}=
    \\&\quad \frac{(-1)^{k+m+s}2^{2 k-q-2 r-2 t+13}(r-t)(-i+k+1) (2 r-1)^{j+1} (2 m-1)^{2 i-2 r+4} (2 t-1)^{q-j}  (2 s-1)^{-2 i+2 k-2 t+6} }{(2 i)! (j)! (-2 i+2 k+2)! (q-j)!}.
    \end{aligned}
	\end{equation}
\end{prop}	

\begin{exmp}
	Some values of $b_{p,q}$ are 
	\begin{equation*}
	\begin{aligned}
		&b_{1,1}=-8,\ b_{1,2}=-2,\ b_{1,3}=-\frac{5}{2},\ b_{1,4}=-\frac{21}{4},\ b_{1,5}=-\frac{231}{16},\ b_{1,6}=-\frac{3003}{64},\ b_{1,7}=-\frac{21879}{128},
        \\&b_{1,8}=-\frac{692835}{1024},\ b_{1,9}=-\frac{11685817}{4096},
		\\& b_{2,2}=2,\ b_{2,3}=16,\ b_{2,4}=168,\ b_{2,5}=2112,\ b_{2,6}=30030,\ b_{2,7}=466752,\\&b_{2,8}=7759752,\ b_{2,9}=135980416,
        \\&b_{3,3}=-\frac{27}{2},\ b_{3,4}=-\frac{2187}{4},\ b_{3,5}=-\frac{649539}{32},\ b_{3,6}=-\frac{98513415}{128},\ b_{3,7}=-\frac{3875799213}{128},
        \\&b_{3,8}=-\frac{1265272270353}{1024},\ b_{3,9}=-\frac{106704628133103}{2048},
		\\&b_{4,4}=384,\ b_{4,5}=45056,\ b_{4,6}=4100096,\ b_{4,7}=347602944,\ b_{4,8}=28894494720,\ b_{4,9}=2399280300032,
		\\&b_{5,5}=-\frac{859375}{32},\ b_{5,6}=-\frac{888671875}{128},\ b_{5,7}=-\frac{161865234375}{128},\\&b_{5,8}=-\frac{207000732421875}{1024},\ b_{5,9}=-\frac{62346649169921875}{2048},
		\\&b_{6,6}=3582306,\ b_{6,7}=1734623424,\ b_{6,8}=556163635320,\ b_{6,9}=150090026385024.
	\end{aligned}
	\end{equation*}
Some values of $d_{p,q}$ are
	\begin{equation*}
	\begin{aligned}
		&d_{1,1}=1,\ d_{1,2}=-\frac{1}{64},\ d_{1,3}=-\frac{21}{512},\ d_{1,4}=-\frac{2145}{8192},\ d_{1,5}=-\frac{323323}{131072},\ d_{1,6}=-\frac{30421755}{1048576},\\&d_{1,7}=-\frac{3305942145}{8388608},
		\ d_{1,8}=-\frac{1590158171745}{268435456},\ d_{1,9}=-\frac{824543781407775}{8589934592},
        \\&d_{2,2}=-\frac{1}{64},\ d_{2,3}=-\frac{567}{1024},\ d_{2,4}=-\frac{312741}{8192},\ d_{2,5}=-\frac{235702467}{65536},\ d_{2,6}=-\frac{427708145475}{1048576},\\&d_{2,7}=-\frac{1756913199480945}{33554432},\ d_{2,8}=-\frac{1971842247550780605}{268435456},\ d_{2,9}=-\frac{2366260407369698510385}{2147483648},
        \\&d_{3,3}=-\frac{525}{1024},\ d_{3,4}=-\frac{1340625}{8192},\ d_{3,5}=-\frac{3608515625}{65536},\ d_{3,6}=-\frac{42441064453125}{2097152},\\&d_{3,7}=-\frac{269038260498046875}{33554432},\ d_{3,8}=-\frac{905851823096923828125}{268435456},\ d_{3,9}=-\frac{3202573724839324951171875}{2147483648},
        \\&d_{4,4}=-\frac{1030029}{8192},\ d_{4,5}=-\frac{38038627627}{262144},\ d_{4,6}=-\frac{292292272742925}{2097152},\ d_{4,7}=-\frac{2178975938039761245}{16777216},
		\\&d_{4,8}=-\frac{32681271562328539043985}{268435456},\ d_{4,9}=-\frac{498218295256742735716364505}{4294967296},
		\\&d_{5,5}=-\frac{24546728349}{262144},\ d_{5,6}=-\frac{561238628492295}{2097152},\ d_{5,7}=-\frac{9431269774195483755}{16777216},
		\\&d_{5,8}=-\frac{285795767967445744227765}{268435456},\ d_{5,9}=-\frac{8310236142024805237932527775}{4294967296},
		\\&d_{6,6}=-\frac{310532064755055}{2097152},\ d_{6,7}=-\frac{28582541707068569715}{33554432},\ d_{6,8}=-\frac{895748274557821906298385}{268435456},
        \\&d_{6,9}=-\frac{24086181828592215443355930675}{2147483648}.
	\end{aligned}
	\end{equation*}	
 Some values of $a_{p,q}$ are 
	\begin{equation*}
	\begin{aligned}
	&a_{0,0}=1,\ a_{0,1}=\frac{1}{2},\ a_{0,2}=\frac{1}{8},\ a_{0,3}=\frac{1}{48},\ a_{0,4}=\frac{1}{384},\ a_{0,5}=\frac{1}{3840},\ a_{0,6}=\frac{1}{46080},\\&a_{0,7}=\frac{1}{645120},\ a_{0,8}=\frac{1}{10321920},\ a_{0,9}=\frac{1}{185794560},
	\\&a_{1,0}=-\frac{1}{2},\ a_{1,1}=-\frac{1}{2},\ a_{1,2}=-\frac{1}{4},\ a_{1,3}=-\frac{1}{12},\ a_{1,4}=-\frac{1}{48},\ a_{1,5}=-\frac{1}{240},\ a_{1,6}=-\frac{1}{1440},\\&a_{1,7}=-\frac{1}{10080},\ a_{1,8}=-\frac{1}{80640},\ a_{1,9}=-\frac{1}{725760},
    \\&a_{2,0}=-\frac{5}{2},\ a_{2,1}=-\frac{7}{4},\ a_{2,2}=-\frac{13}{16},\ a_{2,3}=-\frac{31}{96},\ a_{2,4}=-\frac{85}{768},\ a_{2,5}=-\frac{247}{7680},\ a_{2,6}=-\frac{733}{92160},\\&a_{2,7}=-\frac{313}{184320},\ a_{2,8}=-\frac{1313}{4128768},\ a_{2,9}=-\frac{19687}{371589120},
    \\&a_{3,0}=\frac{1}{3},\ a_{3,1}=-\frac{2}{3},\ a_{3,2}=-\frac{4}{3},\ a_{3,3}=-\frac{10}{9},\ a_{3,4}=-\frac{11}{18},\ a_{3,5}=-\frac{23}{90},\ a_{3,6}=-\frac{47}{540},
    \\&a_{3,7}=-\frac{19}{756},\ a_{3,8}=-\frac{191}{30240},\ a_{3,9}=-\frac{383}{272160},
    \\&a_{4,0}=-\frac{7}{24},\ a_{4,1}=-\frac{179}{48},\ a_{4,2}=-\frac{1199}{192},\ a_{4,3}=-\frac{6779}{1152},\ a_{4,4}=-\frac{36119}{9216},\ a_{4,5}=-\frac{187139}{92160},\\&a_{4,6}=-\frac{955199}{1105920},\ a_{4,7}=-\frac{4834379}{15482880},\ a_{4,8}=-\frac{24346919}{247726080},\ a_{4,9}=-\frac{122259539}{4459069440},
    \\&a_{5,0}=-\frac{76}{15},\ a_{5,1}=-\frac{256}{15},\ a_{5,2}=-\frac{428}{15},\ a_{5,3}=-\frac{1388}{45},\ a_{5,4}=-\frac{1097}{45},\ a_{5,5}=-\frac{3407}{225},\\&a_{5,6}=-\frac{10457}{1350},\ a_{5,7}=-\frac{31847}{9450},\ a_{5,8}=-\frac{96497}{75600},\ a_{5,9}=-\frac{291407}{680400},
    \\&a_{6,0}=-\frac{2401}{144},\ a_{6,1}=-\frac{95467}{1440},\ a_{6,2}=-\frac{738613}{5760},\ a_{6,3}=-\frac{5558491}{34560},\ a_{6,4}=-\frac{8191361}{55296},\ a_{6,5}=-\frac{297250507}{2764800},
    \\&a_{6,6}=-\frac{2134462933}{33177600},\ a_{6,7}=-\frac{15212619451}{464486400},\ a_{6,8}=-\frac{21570744737}{1486356480},\ a_{6,9}=-\frac{761828479147}{133772083200}.
	\end{aligned}
	\end{equation*}	

\end{exmp}	
Note that the solution is a Taylor series of variables $s$ and $t^2-3 {\rm log}(t^3)$.
		
\section*{Acknowledgements}
	This work was done during the author's undergraduate thesis project in 2021-2022. The author would like to thank his thesis advisor Di Yang for suggesting the project and for helpful discussions. 

\printbibliography

\end{document}